\input{aipcheck}

\documentclass[
    ,final            
  ]
  {aipproc}

 \layoutstyle{8x11single}

\usepackage{amsmath}

\newcommand\Ocal{\mathcal{O}}

\newcommand{\ovli}[1]{{\overline{#1}}}

\newcommand\QQbar{Q\overline{Q}}
\newcommand\Qbar{\overline{Q}}
\newcommand\mQ{m_{Q}}
\newcommand\SdotS{{\mathbf S}_Q\cdot{\mathbf S}_{\overline{Q}}}

\newcommand\SQ{{\mathbf S}_Q}
\newcommand\SQbar{{\mathbf S}_{\overline{Q}}}
\newcommand\tsrc{t_{\rm s}}
\newcommand\alt{\hspace{0.3em}\raisebox{0.4ex}{$<$}\hspace{-0.75em}\raisebox{-.7ex}{$\sim$}\hspace{0.3em}}
\newcommand\agt{\hspace{0.3em}\raisebox{0.4ex}{$>$}\hspace{-0.75em}\raisebox{-.7ex}{$\sim$}\hspace{0.3em}}

\begin{document}

\title{Potential description of the charmonium from lattice QCD}

\classification{11.15.Ha, 12.38.-t, 12.38.Gc }
\keywords{Lattice gauge theory, Quantum chromodynamics, Lattice QCD calculations}

\author{Taichi Kawanai}{
  address={J\"ulich Supercomputing Center, J\"ulich D-52425, Germany},
  email={t.kawanai@fz-juelich.de},
  thanks={This work was commissioned by the AIP}
}

\iftrue
\author{Shoichi Sasaki}{
  address={Department of Physics, Tohoku University, Sendai 980-8578, Japan},
  email={ssasaki@phys.s.u-tokyo.ac.jp},
}

 \begin{abstract}
  We present spin-independent and spin-spin interquark potentials for charmonium states, 
  that are calculated using a relativistic heavy quark action for charm quarks
  on the PACS-CS gauge configurations generated with the Iwasaki gauge action and 2+1 flavors of Wilson clover quark.
  The interquark potential with finite quark masses
  is defined through the equal-time Bethe-Salpeter amplitude.
  The light and strange quark masses are close to the physical point where the pion mass corresponds to $M_\pi \approx 156(7)$~MeV, and
  charm quark mass is tuned to reproduce the experimental values of $\eta_c$ and $J/\psi$ states.
  Our simulations are performed with a lattice cutoff of $a^{-1}\approx 2.2$~GeV 
 and a spatial volume of $(3~{\rm fm})^3$. 
  We solve the nonrelativistic Schr\"odinger equation with resulting charmonium potentials 
  as theoretical inputs. The resultant charmonium spectrum below the open charm threshold
  shows a fairly good agreement with experimental data of well-established charmonium states.
 \end{abstract}

\maketitle

\section{\label{intro} Introduction}

The heavy-quark~($Q$)-antiquark~($\Qbar$) potential is 
an important quantity to understand many properties of the heavy quarkonium states.
The dynamics of heavy quarks can be described well 
within the framework of nonrelativistic quantum mechanics, 
because of their masses being much larger than the QCD scale~($\Lambda_{\rm QCD}$).
Indeed the constituent quark potential models with a QCD-motivated $\QQbar$ potential have 
successfully reproduced the heavy quarkonium spectra and also decay rates 
below open thresholds~\cite{Eichten:1974af, Godfrey:1985xj, Barnes:2005pb}.

In the nonrelativistic potential~(NRp) models,
the heavy quarkonium states such as charmonium and bottomnium 
are well understood as is a quark-antiquark pair bound by the Coulombic
induced by perturbative one-gluon exchange, plus linearly rising potential.
The former dominates in short range, while the latter describes the phenomenology of 
confining quark interactions at large distances~\cite{Eichten:1974af}.
This potential is called as Cornell potential and its functional form is given by 
$V(r) = - \frac{4}{3} \frac{\alpha_s}{r} + \sigma r + V_0$
where $\alpha_s$ and $\sigma$ denote the strong coupling constant and the string tension, 
and $V_0$ is the constant term associated with a self-energy contribution of the color sources.
In the NRp models, spin-dependent potentials are induced as relativistic corrections
in powers of the relative velocity of quarks, and their functional forms are also
determined on the basis of perturbative one-gluon exchange as the Fermi-Breit type potential~\cite{Eichten:1980mw}.
However the validity of the phenomenological spin-dependent potentials determined within the  perturbative method 
would be limited  only at short distances and also in the vicinity of the heavy quark mass limit.
This may cause large uncertainties in the predictions for higher heavy quarkonium states obtained in the NRp models.

Lattice QCD simulations offer a strong tool to understand the properties of $\QQbar$ interactions.
Indeed, both the static $\QQbar$ potential and its corrections of order $\Ocal(1/m_Q^2)$ as the spin dependent potentials
have been precisely determined from Wilson loops using lattice QCD simulations 
with the multilevel algorithm~\cite{Koma:2006fw,Koma:2010zz}.
Although the lattice QCD calculations within the Wilson loop formalism 
support a shape of the Cornell potential~\cite{Bali:2000gf},
the leading spin-spin potential determined at $\Ocal(1/m_Q^2)$ gives
an attractive interaction for the higher spin states~\cite{Bali:1996cj, Bali:1997am},
in contradiction with a repulsive one that is demanded by phenomenological analysis.
The higher order corrections beyond the next-to-leading order are 
required to correctly describe the conventional charmonium spectrum,
because the inverse of the charm quark mass would be far outside the validity region of the $1/\mQ$ expansion~\cite{Kawanai:2013aca}. 
In addition, practically, the multilevel algorithm is quite difficult 
to be implemented in dynamical lattice QCD simulations.

Under this situation, we employ the new method proposed 
in our previous works~\cite{Kawanai:2013aca, Kawanai:2011xb, Kawanai:2011jt}
in order to obtain the proper interquark potentials fitted in the NRp models.
The interquark potential and also the quark kinetic mass are defined by the equal-time and 
Coulomb gauge Bethe-Salpeter~(BS) amplitude through an effective Schr\"odinger equation.
This new method enables us to determine the interquark potentials including spin-dependent terms
at {\it finite quark masses} from first principles of QCD, and then can fix all parameters 
that are needed in the NRp models.
Furthermore, there is no restriction to extend to the dynamical calculations.
Hereafter we call the new method as {\it BS amplitude method}.

Once we obtain the reliable $\QQbar$ potentials from lattice QCD,
we can solve the nonrelativistic Schr\"odinger equation 
with ``lattice-determined potentials'' as theoretical inputs, and obtain many physical observables such as mass spectrum.
In this proceedings, we present not only the charmonium potentials 
calculated with almost physical quark masses using the $2+1$ flavor PACS-CS gauge configurations~\cite{Aoki:2008sm},
but also the resultant charmonium mass spectrum computed from the NRp model
with the lattice-determined potentials, where there are no free parameters including the $V_0$ and quark mass. 
The simulated pion mass~$M_\pi \approx 156(7)$~MeV is almost physical.
For the heavy quarks, we employ the relativistic heavy quark action that 
can control large discretization errors introduced by large quark mass~\cite{Aoki:2001ra}.

%
%
\section{\label{formalism} Formalism}
In this section, we briefly review the BS amplitude  method utilized to
calculate the interquark potential with the finite quark mass.
This is an application based on 
the approach originally used for studing the hadron-hadron potential, 
which is defined through the equal-time BS amplitude~\cite{Ishii:2006ec,Aoki:2009ji}.
More details of determination of the interquark potential are given in Ref.~\cite{Kawanai:2013aca}.

In lattice simulations,
we measure the following equal-time $\QQbar$ 
BS amplitude in the Coulomb gauge for the 
quarkonium states~\cite{{Velikson:1984qw},{Gupta:1993vp}}:
\begin{equation}
  \phi_\Gamma({\bf r})= \sum_{{\bf x}}\langle 0| \overline{Q}
  ({\bf x})\Gamma Q({\bf x}+{\bf r})|
  \QQbar;J^{PC}\rangle \label{eq_phi},
\end{equation}
where ${\bf r}$ is the relative coordinate between quark and antiquark at time slice $t$.
The Dirac $\gamma$ matrices $\Gamma$ in Eq.~(\ref{eq_phi}) specify the spin and the parity of meson states.
For instance, $\gamma_5$ and $\gamma_i$ correspond to 
the pseudoscalar (PS)  and the vector (V)  channels
with $J^{PC}=0^{-+}$ and $J^{PC}=1^{--}$, respectively.
A summation over spatial coordinates ${\bf x}$ projects onto zero total momentum. 
The ${\bf r}$-dependent amplitude, $\phi_\Gamma({\bf r})$, is here called {\it BS wave function}.
The BS wave function can be extracted from four-point correlation function
at large time separation.
Also, the corresponding meson masses $M_\Gamma$ can be read off from the asymptotic large-time behavior 
of two-point correlation functions.
In this proceedings, we focus only on the $S$-wave charmonium states ($\eta_c$ and $J/\psi$),
obtained by appropriate projection to the $A^{+}_{1}$ representation in cubic group~\cite{Luscher:1990ux}.

The BS wave function satisfies an effective Schr\"odinger
equation with a nonlocal and energy-independent 
interquark potential $U$~\cite{Ishii:2006ec,Caswell:1978mt,Ikeda:2011bs} 
\begin{equation}
  -\frac{\nabla^2}{2\mu}\phi_\Gamma({\bf r})+
  \int dr'U({\bf r},{\bf r}')\phi_\Gamma({\bf r}')
  =E_\Gamma\phi_\Gamma({\bf r}),
  \label{Eq_schr}
\end{equation}
where $\mu$ is the reduced mass of the $\QQbar$ system.
The energy eigenvalue $E_\Gamma$ of the stationary 
Schr\"odinger equation is supposed to be $M_\Gamma-2\mQ$.
If the relative quark velocity $v=|{\nabla}/\mQ|$ is small as $v \ll 1$, 
the nonlocal potential $U$ can generally expand in terms of the velocity $v$ as 
$U({\bf r}',{\bf r})=  \{V(r)+V_{\text{S}}(r)\SdotS +V_{\text{T}}(r)S_{12}+
 V_{\text{LS}}(r){\bf L}\cdot{\bf S} + \mathcal{O}(v^2)\}\delta({\bf r}'-{\bf r})$
where $S_{12}=(\SQ\cdot\hat{r})(\SQbar\cdot\hat{r})-\SdotS/3$
with $\hat{r}={\bf r}/r$, ${\bf S}=\SQ+\SQbar$
and ${\bf L} = {\bf r}\times (-i\nabla)$~\cite{Ishii:2006ec}.
Here, $V$, $V_{\text{S}}$, $V_{\text{T}}$ and $V_{\text{LS}}$ represent
the spin-independent central, spin-spin, tensor and spin-orbit potentials, 
respectively.
 
The Schr\"odinger equation for $S$-wave is simplified as 
\begin{equation}
  \left\{
  - \frac{\nabla^2}{\mQ}
  +V(r)+\SdotS V_{\text{S}}(r)
  \right\}\phi_{\Gamma}(r)=E_\Gamma \phi_{\Gamma}(r)
  \label{Eq_pot}
\end{equation}
at the leading order of the $v$-expansion.
Here, we essentially follow the NRp models,
where the $J/\psi$ state is purely composed of the $1S$ wave function. 

The spin operator $\SdotS$ can be easily replaced by expectation values 
$-3/4$ and $1/4$ for the PS and V channels, respectively. 
Then, the spin-independent and spin-spin $\QQbar$ potentials can be  evaluated 
through the following linear combinations of Eq.(\ref{Eq_pot}):
 \begin{eqnarray}
   V(r)
   &=& E_{\text{ave}}+\frac{1}{\mQ}\left\{
   \frac{3}{4}\frac{\nabla^2\phi_\text{V}(r)}{\phi_\text{V}(r)}+
    \frac{1}{4}\frac{\nabla^2\phi_\text{PS}(r)}{\phi_\text{PS}(r)}
   \right\} \label{Eq_potC}\\
   V_{\text{S}}(r) 
   &=& E_{\text{hyp}} + \frac{1}{\mQ}\left\{
  \frac{\nabla^2\phi_\text{V}(r)}{\phi_\text{V}(r)} 
  - \frac{\nabla^2\phi_\text{PS}(r)}{\phi_\text{PS}(r)} \right\},\label{Eq_potS}
 \end{eqnarray}
 where $E_{\text{ave}}=M_{\text{ave}}-2\mQ$ 
 and $E_{\text{hyp}}=M_\text{V}-M_\text{PS}$.
 The mass $M_{\text{ave}}$ denotes the spin-averaged mass as  
 $\frac{1}{4}M_\text{PS}+\frac{3}{4}M_\text{V}$.
 The derivative $\nabla^2$ is defined by the discrete Laplacian.

The kinetic quark mass is an important quantity in the determination 
of the interquark potentials
since Eqs.~(\ref{Eq_potC}) and (\ref{Eq_potS})
require an information of the kinetic quark mass $m_Q$.
In our previous work~\cite{Kawanai:2013aca,Kawanai:2011xb,Kawanai:2011jt},
we propose to calculate the quark kinetic mass through the large-distance behavior 
in the spin-spin potential with the help of the measured hyperfine splitting energy of
$1S$ states in heavy quarkonia.
Under a simple, but reasonable assumption as $\lim_{r\to\infty} V_S(r)=0$  
which implies there is no long-range correlation and no irrelevant constant term 
in the spin-spin potential, 
Eq.~(\ref{Eq_potS}) is rewritten as 
\begin{equation}
 m_Q = \lim_{r\to \infty} \frac{-1}{E_\text{hyp}} \left\{ \frac{\nabla^2\phi_\text{V}(r)}{\phi_\text{V}(r)} 
  - \frac{\nabla^2\phi_\text{PS}(r)}{\phi_\text{PS}(r)} \right\},
\label{eq_quark_mass}
\end{equation}
and then  we can estimate the kinetic quark mass from asymptotic behavior of Eq.~(\ref{eq_quark_mass}) in long range region.

\section{\label{setup}Lattice setup}
%

%
%
\begin{table}
\caption{Parameters of $2+1$-flavor dynamical QCD gauge field configurations
generated by PACS-CS collaboration~\cite{Aoki:2008sm}.
The columns list number of flavors, lattice volume, the $\beta$ value,
hopping parameters (light, strange), approximate lattice spacing~(lattice cut-off),
spatial physical volume, pion mass,  number of configurations to be analyzed. }
\label{tab:ensembles_full}
  \begin{tabular}{ccccccccc}
   \hline 
  $N_f$ & $L^3\times T$   &$\beta$ & $\kappa_{ud}$ &$\kappa_{s}$ &$a$~[fm]~($a^{-1}$ [GeV])    
  & $La$~[fm]  & $M_\pi$~[MeV] & \# configs. \\[2pt] \hline
  $2+1$ &$32^3\times 64$  &1.9     & 0.13781 & 0.13640 & $\approx$ 0.0907~($\approx$ 2.176)  & $\approx$ 2.90      & $\approx$156 & 198\\  \hline
  \end{tabular}
\end{table}

The computation of the charmonium potential in this study is performed on
a lattice $L^3\times T=32^3\times 64$ using the $2+1$ flavor PACS-CS gauge configurations~\cite{Aoki:2008sm}
generated
by non-perturbatively $\mathcal{O}(a)$-improved Wilson quark action with $c_{SW} = 1.715$~\cite{Aoki:2005et}
and Iwasaki gauge action at $\beta=1.90$~\cite{Iwasaki:2011np},
which corresponds to a lattice cutoff of $a^{-1} =  2.176(31)$ GeV ($a = 0.0907(13)$fm).
The spatial lattice size then corresponds to $La \approx 3\;{\rm fm}$.
The hopping parameters for the light sea quarks \{$\kappa_{ud}$,$\kappa_{s}$\}=\{0.13781, 0.13640\}
provide  $M_\pi=156(7)$ MeV and $M_K= 554(2)$ MeV~\cite{Aoki:2008sm}.
Table~\ref{tab:ensembles_full} summarizes simulation parameters of dynamical QCD simulations
used in this work.
Although the light sea quark masse is slightly off the physical point,
the systematic uncertainty due to this fact could be extremely small
in this project.
Our results are analyzed on all 198 gauge configurations.
All gauge configurations are fixed to Coulomb gauge.

%
%
\begin{table}[t]
  \caption{
    The hopping parameter $\kappa_Q$ and RHQ parameters used for the charm quark. 
      \label{tab:RHQ_para_charmonium}
      }
      \begin{tabular}{ccccc} \hline
	$\kappa_c$ & $\nu$ & $r_s$  & $c_B$ & $c_E$  \\ \hline
	0.10819 & 1.2153 & 1.2131  & 2.0268 & 1.7911  \\ \hline
      \end{tabular} 
\end{table}
In order to control discretization errors induced by large quark mass,
we employ the relativistic heavy quark~(RHQ) action~\cite{Aoki:2001ra} 
that removes main errors of $\Ocal(|\vec{p}|a)$, $\Ocal((m_0 a)^n )$ and $\Ocal(|\vec{p}|a (m_0 a)^n)$
from on-shell Green's functions.
The RHQ action is the anisotropic version of the $\Ocal (a)$ improved Wilson action 
with five parameters $\kappa_c$, $\nu$, $r_s$, $c_B$ and $c_E$,
called {\it RHQ parameters} (for more details see Ref.~\cite{Aoki:2001ra,Kayaba:2006cg}).
The RHQ action utilized here is a variant of the Fermilab
approach~\cite{ElKhadra:1996mp}~(See also Ref.~\cite{Christ:2006us}).

The parameters $r_s$, $c_B$ and $c_E$ in RHQ action are determined
by tadpole improved one-loop perturbation theory~\cite{Kayaba:2006cg}.
For $\nu$, we use a nonperturbatively determined value,
which is adjusted by reproducing the effective speed of light $c_{\text{eff}}$
to be unity in the dispersion relation
$E^2({\bf p}^2)=M^2+c^2_{\text{eff}}|{\bf p}|^2$
for the spin-averaged $1S$-charmonium state, since the parameter $\nu$
is sensitive to the size of hyperfine splitting energy~\cite{Namekawa:2011wt}.
We choose $\kappa_c$ to reproduce the experimental spin-averaged mass
of $1S$-charmonium states $M_\text{ave}^{\text{exp}}(1S)=3.0678(3)$~GeV.
To calibrate adequate RHQ parameters, we employ a gauge invariant
Gauss smearing source for the standard two-point correlation function
with four finite momenta.
As a result, the relevant speed of light in the 
dispersion relation is consistent with unity within statistical error: $c^2_{\text{eff}}=1.04(5)$.
Our chosen RHQ parameters are summarized in Table~\ref{tab:RHQ_para_charmonium}.

Using tuned RHQ parameters, 
we compute two valence quark propagators with wall sources located
at different time slices $\tsrc/a=6$ and $57$ to increase statistics.
Two sets of two and four-point correlation functions 
are constructed from the corresponding quark propagators, and 
folded together to create the single correlation function.
Dirichlet boundary condition is imposed for the time direction
to eliminate unwanted contributions across time boundaries.

%
\begin{table}
      \begin{tabular}{rccc} \hline
       state~($J^{PC}$) & fit range & mass~[GeV] & $\chi^2/{\rm d.o.f.}$\\ \hline
      $\eta_c$ ($0^{-+}$)     & [33:47] & 2.9851(5)  & 0.70\\
      $J/\psi$ ($1^{-+}$)     & [33:47] & 3.0985(11) & 0.62\\ 
      $\chi_{c0}$ ($0^{++}$)  & [14:26] & 3.3928(59) & 0.66\\  
      $\chi_{c1}$ ($1^{++}$)  & [14:26] & 3.4845(62) & 1.03\\ 
      $h_{c}$ ($1^{+-}$)      & [14:26] & 3.5059(62) & 0.63\\[2pt] 
      $M_{\text{ave}}(1S)$    & - & 3.0701(9)  & - \\ 
      $E_{\text{hyp}}(1S)$    & - & 0.1138(8)  & - \\ \hline
      \end{tabular}

 \caption{
    Masses of low-lying charmonium states calculated from two-point functions, 
    the spin-averaged mass and hyperfine splitting energy of $1S$ charmonium states.
    The fitting ranges and values of $\chi^2/{\rm d.o.f.}$ are also included.
 Results are shown in units of GeV.
  }
    \label{tab:charmonium_mass}
\end{table}

Low-lying  charmonium masses of $\eta_c$, $J/\psi$, $h_c$, $\chi_{c0}$ and $\chi_{c1}$ are obtained by
weighted average of the effective mass in the appropriate range. The effective mass is defined as 
\begin{equation}
 M_\Gamma(t) = \log\frac{G_\Gamma(t,\tsrc)}{G_\Gamma(t+1,\tsrc)}, 
\end{equation}
where $G_\Gamma(t,\tsrc)$ is a  two-point function obtained by setting  ${\bf r}$ to be zero 
in  four-point function $G_\Gamma({\bf r}, t,\tsrc)$.
In Table~\ref{tab:RHQ_para_charmonium}, we summarize resultant charmonium masses 
together with fit ranges used in the fits and $\chi^2/{\rm d.o.f.}$ values.
We take into account a correlation between effective masses measured at various time slices in the fit.
The statistical errors are estimated by the jackknife method.

Low-lying charmonium masses calculated  in this study below $D\bar{D}$ threshold
are all close to the experimental values, though
the hyperfine mass splitting $M_\text{hyp}=0.1124(9)$ GeV
is slightly smaller than the experimental value,
$M_\text{hyp}^{\text{exp}}=0.1166(12)$ GeV~\cite{Beringer:1900zz}.
Note that here we simply neglect the disconnected diagrams in two-point correlation functions.
The several numerical studies reported that the contributions of charm annihilation to
the hyperfine splitting of the $1S$-charmonium state are
sufficiently small, as of order $1-4$~MeV.~\cite{McNeile:2004wu,deForcrand:2004ia,Levkova:2010ft},

\section{\label{main} Determination of interquark potential}
\subsection{\label{wavefunc} $\QQbar$ BS wave function}
 \begin{figure}
   \centering
   \includegraphics[width=.49\textwidth]{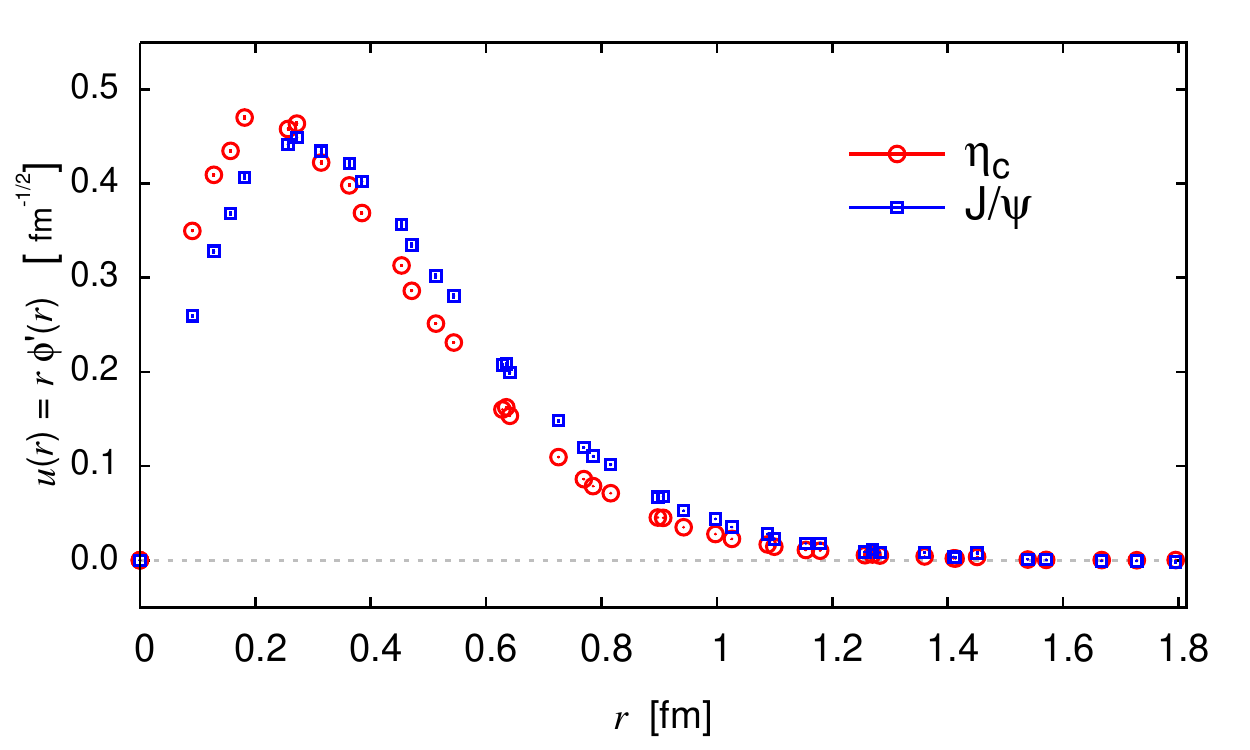}
   \caption{ The reduced $\QQbar$ BS wave functions of the $\eta_c$~(circles)
    and $J/\psi$~(squares) states,  shown as a function of the spatial distance~$r$.
    The data points are taken along $\bf r$ vectors which are multiples of 
   three directions~$(1,0,0)$, $(1,1,0)$ and $(1,1,1)$.
   }
    \label{wavefunction1}
 \end{figure}

Fig.~\ref{wavefunction1} shows the  $\QQbar$ BS wave functions of
$1S$ charmonium states ($\eta_c$ and $J/\psi$ states).
The BS wave functions are defined by Eq.(\ref{eq_phi}) and normalized as $\sum \phi_{\Gamma}^2 = 1$.
We use the reduced wave function $u_\Gamma(r)$ for displaying the wave function: $u_\Gamma(r) =r \phi_\Gamma({\bf r})$.
Practically we take average of the BS wave function by weight over time slices $33 \leq t/a \leq 47$
where effective mass plots for $1S$-charmonium states show plateaus and
excited state contaminations are expected to be negligible.
In Fig.~\ref{wavefunction1}  we display data points of $u_\Gamma(r)$ 
calculated at $\bf r$ vectors which are multiples of $(1,0,0)$, $(1,1,0)$ and $(1,1,1)$.
Hereafter we focus on lattice data taken in three directions for any quantities.

We find that a sign of rotational symmetry breading found in
the $\QQbar$ BS wave functions is sufficiently small in our calculation.
The resulting  wave functions become isotropic
with the help of a projection to the $A_1^+$ sector of the cubic group
that corresponds to the $S$-wave in the continuum theory~(Fig.~\ref{wavefunction1}).

\subsection{\label{determination_quarkmass}quark kinetic mass}
 \begin{figure}
   \centering
   \includegraphics[width=.49\textwidth]{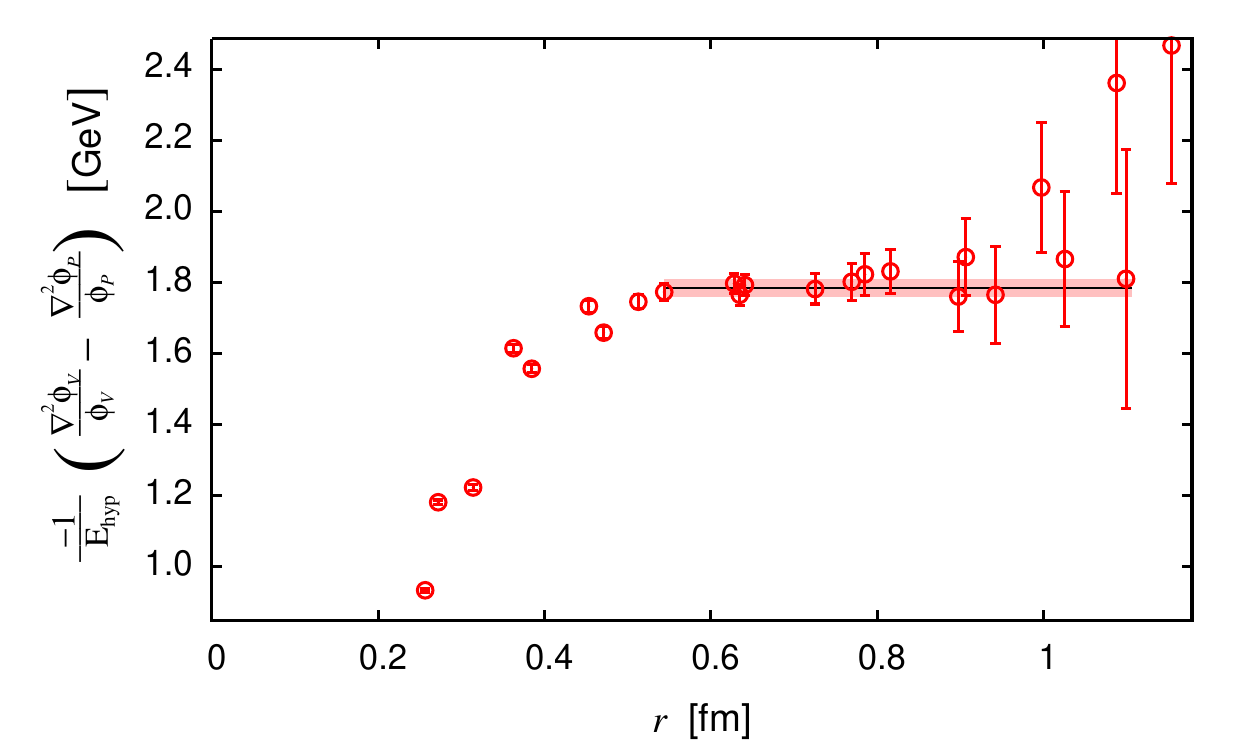}
   \caption{
    The determination of quark kinetic mass within the BS amplitude method.
    The values of $-(\nabla^2\phi_V/\phi_V-\nabla^2\phi_P/\phi_P)/E_\text{hyp}$
    as a function of the spatial distance $r$ are shown in this figure.
    The quark kinetic mass $m_Q$ is obtained from the long-distance asymptotic values of
    $-(\nabla^2\phi_V/\phi_V-\nabla^2\phi_P/\phi_P)/E_\text{hyp}$.
    Horizontal solid line indicates a value of quark kinetic mass
    obtained by fitting a asymptotic constant in the range $0.54~\text{fm}\alt r \alt 1.10~\text{fm}$.
    A shaded band indicates a statistical error estimated by estimated by jackknife method.
  }
  \label{fig:quarkmass_charm}
 \end{figure}
In our formalism, the kinetic mass of the charm quark is
determined self-consistently within the BS amplitude method as well~\cite{Kawanai:2011xb}.
The quark kinetic mass defined in Eq.~(\ref{eq_quark_mass}) is calculated
from asymptotic behavior of the quantity $-(\nabla^2\phi_V/\phi_V-\nabla^2\phi_P/\phi_P)/E_\text{hyp}$  at long distances.
Fig.~\ref{fig:quarkmass_charm} illustrates the determination of quark kinetic mass $m_Q$ for the charmonium system.

For the derivative, we use the discrete Laplacian operator $\nabla^2$ defined in polar coordinates as
\begin{eqnarray}
\nabla_{\bf r}^2 \phi_\Gamma(r)
= \frac{2}{r}\frac{\phi_\Gamma(r+\tilde{a})-\phi_\Gamma(r-\tilde{a})}{2\tilde{a}} \nonumber 
 + \frac{\phi_\Gamma(r+\tilde{a})+\phi_\Gamma(r-\tilde{a})-2\phi_\Gamma}{\tilde{a}^2} 
\label{eq_laplacian_polar}
\end{eqnarray}
where $r$ is the absolute value of the relative distance as $r = |{\bf r}|$ and
$\tilde{a}$ is a spacing between grid points along differentiate directions. 
In the on-axis~$( {\bf r} \propto (1,0,0) )$ and the two off-axis directions (${\bf r}\propto  (1,1,0)$ and $(1,1,1)$),
the effective grid spacings correspond to $\tilde{a} = a, \sqrt{2}a, \sqrt{3}a$, respectively.

The differences of ratios $\nabla^2 \phi_\Gamma /\phi_\Gamma$ at each ${\bf r}$ are obtained by a constant fit to the lattice data  
with a reasonable $\chi^2/{\rm d.o.f.}$ value
over the range of time slices where two-point functions exhibit the plateau behavior~($33 \leq t/a \leq 47$).
Then the values of $m_Q$ are determined for each directions from asymptotic values of 
 $-(\nabla^2\phi_V/\phi_V-\nabla^2\phi_P/\phi_P)/E_\text{hyp}$
in the range of  $ 6 \leq r/a \leq 7\sqrt{3}$ where $V_S(r)$ should vanish.
Finally we average them over three directions, and then obtain $m_Q = 1.784(23)(6)(20)$~GeV.
The first error is statistical, given by the jackknife analysis.
In the second error, we quote a systematic uncertainty due to rotational symmetry breaking
by taking the largest difference between the average value and individual ones obtained for specific directions.
The third one represents the systematic uncertainties due to choice of $t_{\rm min}$ of the time range used in the fits.
We vary $t_{\rm min}$ over range $33-41$ and then quote the largest difference from the preferred determination of $\mQ$.

\subsection{\label{sec:potential}Spin-independent interquark potential}
 \begin{figure}
   \centering
   \includegraphics[width=.49\textwidth]{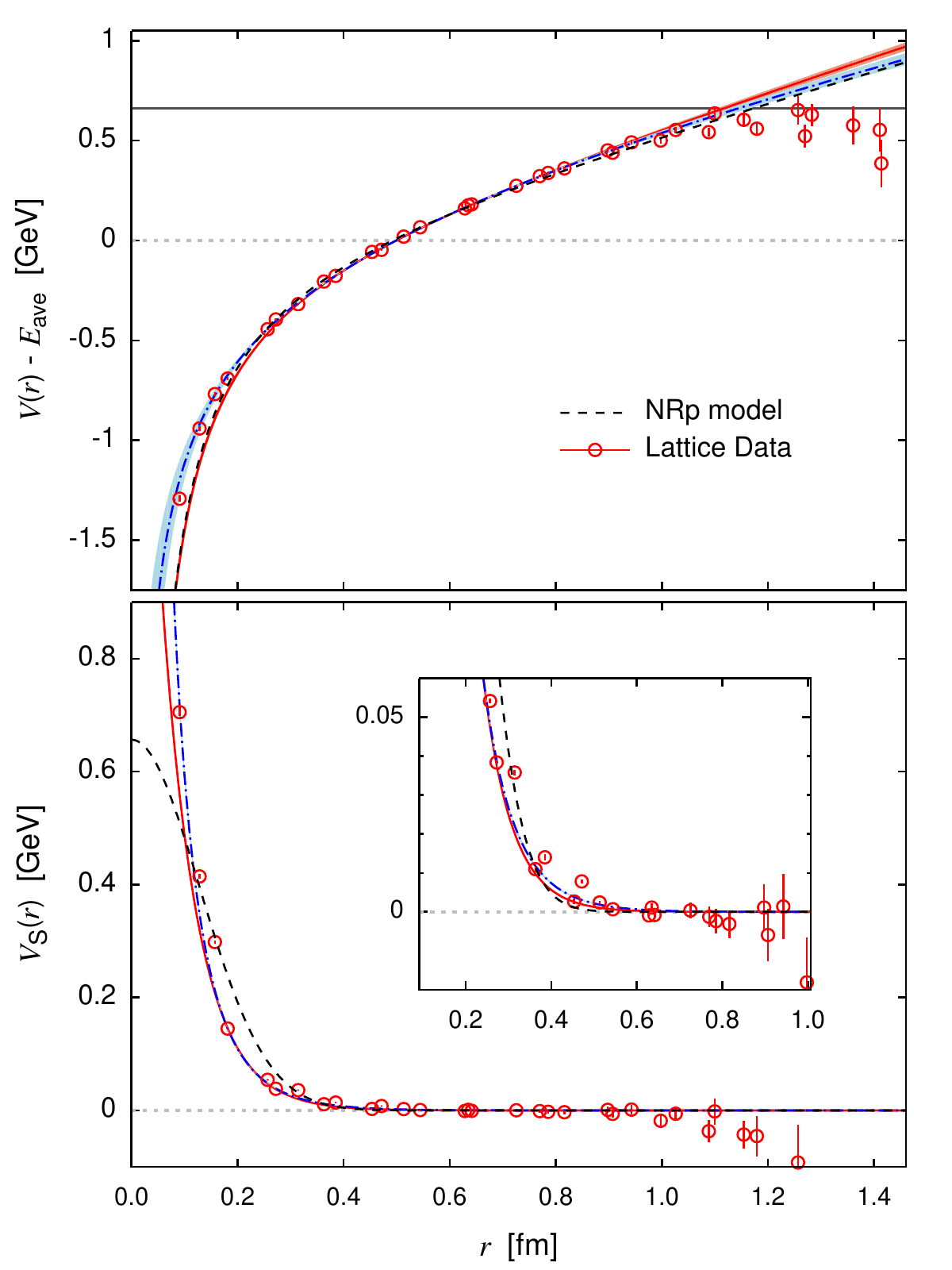}
  \caption{ 
    Central spin-independent and spin-spin charmonium potentials
    calculated from the BS wave functions in the dynamical QCD simulation
    with almost physical quark masses.
    In the upper panel, we show the spin-independent potential $V(r)$.
    A solid (dot-dashed) curve is the fit results
    with the Cornell~(Cornell plus log) form.
    The shaded bands show statistical uncertainties
    in the fitting procedure where the jackknife analysis is used.
    Note that the spin-averaged eigen-energy of $1S$-charmonium state $E_{\rm ave}$
    is not subtracted in this figure.          
    A horizontal line indicates the level of open-charm~($D^0 \bar{D}^0$) threshold $\approx 3729$~MeV.                                
    In the lower panel, we show the spin-spin potential $V_S(r)$.
    A solid~(dot-dashed) curve corresponds to fitting results with exponential~(Yukawa) form.
    The inset shows a magnified view.
    In both plots,  the phenomenological potentials adopted in a NRp model~\cite{Barnes:2005pb}
    are also included as  dashed curves for comparison.
    \label{fig:potential_charm}
  }
 \end{figure}
%
 \begin{table}
  \centering
   \caption{Summary of the Cornell parameters and the quark mass
     determined by the BS amplitude method. 
     For comparison, 
     ones adopted in a phenomenological NRp model~\cite{Barnes:2005pb}
    and  ones of the static potential obtained from Polyakov line correlations
    are also included.
    In the first column, the quoted errors indicate the sum of the statistical and
    systematic added in quadrature.    
     \label{tab:parameter_charm}
   }
     \begin{tabular}{cccccc} \hline
       & This work& Polyakov lines & NRp model \\[2pt]        \hline
       $A$                   & 0.713(83)  & 0.476(81)   &0.7281 \\
       $\sqrt{\sigma}$ [GeV] & 0.402(15)  & 0.448(16)   &0.3775 \\
       $\mQ$ [GeV]           & 1.784(31)  & $\infty$   &1.4794 \\ \hline
     \end{tabular}
 \end{table}

Once the quark kinetic mass is determined,
we can easily calculate the central spin-independent and spin-spin
charmonium potentials from the $Q\bar{Q}$ BS wave function
through Eqs.~(\ref{Eq_potC}) and (\ref{Eq_potS}).
First, we show a result of the spin-independent charmonium potential $V(r)$
in Fig.~\ref{fig:potential_charm}.
The constant energy shift $E_{\rm ave}$ is not subtracted.
At each distance ${\bf r}$, the values of interquark potentials $V(r)$ and $V_S(r)$ are practically determined 
by constant fits to data points over time slices where two-point functions exhibit the plateau behavior.
The correlations between data points at different time slices are taken into account in the fitting process.

The charmonium potential calculated by the BS amplitude
method from dynamical attice QCD simulations
properly exhibits the linearly rising potential
at large distances and the Coulomb-like potential at short distances.
The finite $\mQ$ corrections could be  encoded into the Cornell parameters,
although the charm quark mass region would be beyond the radius of convergence 
for the systematic $1/\mQ$ expansion.
Therefore, as first step, we simply adopt the Cornell parametrization to fit the data of 
the spin-independent central potential: $V(r)=-\frac{A}{r}+\sigma r+V_0$
with the Coulombic coefficient $A$, the string tension $\sigma$, and a constant $V_0$.

All fits are performed individually for each three directions over the range 
$[r_{\rm min}/a, r_{\rm max}/a] = [4: 7\sqrt{3}]$. 
We minimize the $\chi^2/{\rm d.o.f}$ including the covariance matrix.
Resulting Cornell parameters of the charmonium potential are
$A=0.713(26)(38)(31)(62)$ and $\sqrt{\sigma}=0.402(6)(4)(9)(9)$ MeV
with $\chi^2/\text{d.o.f.} \approx 3.2$. 
The first error is statistical and the second, third and forth ones are systematic uncertainties 
due to the choice of the differentiate direction, $t_{\rm min}$ and $r_{\min}$, respectively.
The resulting Cornell parameters are summarized in Table~\ref{tab:parameter_charm}.
Also we include both phenomenological ones adopted in the NRp model~\cite{Barnes:2005pb}
and of the static potential obtained from Polyakov loop correlations.
The latter is calculated using the same method as in Ref.~\cite{Aoki:2008sm}.
%
Additionally we calculate the Sommer parameter defined as $r_0 = \sqrt{(1.65 -A)/\sigma}$, 
and then obtain $r_0 = 0.476(6)(11)(3)(6)$~fm, 
which is fairly consistent with the value quoted in Ref.~\cite{Aoki:2008sm}.

As shown in Table~\ref{tab:parameter_charm}, a gap for the Cornell parameters between
the conventional static potential from Wilson-loops~(Polyakov-loops) and
the phenomenological potential used in the NRp models seems to be filled by
our new approach, which nonperturbatively accounts for a finite quark mass effect.
In the charmonium potential from the BS wave function, a Coulomb-like behavior is  enhanced
and the linearly rising force is slightly reduced due to finite charm quark mass effects.
For the spin-independent central interquark potential,
the $1/m_Q$ expansion within the Wilson-loop approach converges
in the heavy quark mass region of $m_Q \agt 1.8$~GeV.
Indeed, as reported in Ref.~\cite{Laschka:2012cf},
the static $\QQbar$ potential and its $1/\mQ$ corrections calculated in Ref.~\cite{Koma:2006si}
agree with the charmonium potential obtained from the BS amplitude method.

In order to provide a more adequate fit to the lattice data, 
we try to employ an alternative functional form adding a $\log$ term to the Cornell potential:
\begin{equation}
  V(r)=-\frac{A}{r}+\sigma r+V_0 + B \log(r\Lambda) \label{eq_Cornell}
\end{equation}
where $\Lambda$ is simply set to be lattice cutoff $a^{-1}$.
Such $\log$ term as $1/\mQ$ corrections to the spin-independent potential 
is reported in Ref.~\cite{Koma:2009ws}.
Resulting parameters are 
$A= 0.194(137)(33)(36)(66)$, $\sqrt{\sigma} = 0.300(38)(19)(20)(21)$~GeV and $B = 0.390(113)(20)(39)(61)$~GeV.
with $\chi^2/\text{d.o.f.} \approx 2.3$.
Fitting range is determined to minimized a $\chi^2/\text{d.o.f.}$ value 
taking into account the correlation, and then we choose $[r_{\rm min}/a, r_{\rm max}/a] = [3: 7\sqrt{3}]$.

The finite quark mass corrections to spin-independent potential 
give only a minor modification in the NRp models.
In the upper panel of Fig.~\ref{fig:potential_charm}
the solid (dot-dashed) curve is given by the fitting the data to the Cornell form (Cornell plus log form).
The phenomenological potential used in the NRp models~\cite{Barnes:2005pb}
is also plotted as a dashed curve for comparison.
The charmonium potential obtained from
lattice QCD is similar to the one used in the NRp models, 
although a slope of the charmonium potential in the long range is barely larger 
than the phenomenological one.  

It is worth mentioning that a {\it string breaking}-like behavior found in the range $r\alt 1.1$ fm is unreliable.
In principle, string breaking due to the presence of dynamical quarks is likely to be observed.
The signal-to-noise ratio however becomes worse rapidly for 
the spin-independent potential as spatial distance $r$ increase because of the localized wave function.
The lattice data of the potential near the spatial boundary are also sensitive to finite volume effects.
Therefore, at least, calculations of the higher charmonium
near the open charm threshold using a larger lattice 
 are required for observing the string breaking.
Their wave functions are extended until the string breaking sets in.

\subsection{\label{sec:potential}Spin-Spin potential}
%
 \begin{table}
  \centering
   \caption{
    Results of fitted parameters for the spin-spin potential with the exponential and Yukawa forms.
    The quoted errors are statistical only. 
    In the case of the spin-spin potential, we use only on-axis data.
     \label{tab:parameter_charm2}
   }
     \begin{tabular}{cccccc} \hline
      Functional form & $\alpha$ & $\beta$ &  $\chi^2/\text{d.o.f.}$ \\[2pt]        \hline
      Exponential     &  2.15(7)~GeV         & 2.93(3)~GeV & 2.0\\
      Yukawa          &  0.815(27)\ \ \ \ \  & 1.97(3)~GeV & 1.7\\ \hline
     \end{tabular}
 \end{table}
 The lower plot of Fig.~\ref{fig:potential_charm} shows the spin-spin
 charmonium potential obtained from the BS amplitude method with almost physical quark masses.
 The spin-spin potential exhibits the short-range {\it repulsive interaction},
 which is required to leads heavier mass to the higher spin state in hyperfine multiplets.
 In contrast of the case of the spin-independent potential,
 the spin-spin potential obtained from BS wavefunction is absolutely different
 from a repulsive $\delta$-function potential generated
 by perturbative one-gluon exchange~\cite{Eichten:1980mw}.
 Such contact form  $\propto \delta({\bf r})$ of the Fermi-Breit type potential
 is widely adopted in the NRp models~\cite{Godfrey:1985xj}.

 The $Q\bar{Q}$ interaction is not entirely due to one-gluon exchange 
 so that spin-spin potential is not necessary to be a simple contact form
 $\propto \delta({\bf r})$. 
 Indeed, the finite-range spin-spin potential described
 by the Gaussian form is adopted by the phenomenological NRp model in Ref.~\cite{Barnes:2005pb}, where
 many properties of conventional charmonium states at higher masses are predicted.
 This phenomenological spin-spin potential is also plotted
 in the lower plot of Fig.~\ref{fig:potential_charm} for comparison.
 There is a slight difference at very short distances,
 although the range of spin-spin potential calculated from the BS amplitude method 
 is similar  to the phenomenological one.

  To examine an appropriate functional form for the spin-spin potential,
  we try to fit the data with several functional forms,
  and explore which functional form can give a reasonable fit
  over the range of $r/a$ from $2$ to $7\sqrt{3}$.  
  As a results, the long-range screening observed in the spin-spin potential
  is accommodated by the exponential form or the Yukawa form:
  \begin{equation}
    V_{\text{S}}(r)=\left\{
    \begin{array}{lcl}
      \alpha \exp(-\beta r)   &:& \text{Exponential form} \\
      \alpha \exp(-\beta r)/r &:& \text{Yukawa form} 
    \end{array}
    \right.
  \end{equation}
  All results of correlated $\chi^2$ fits are summarized in Table~\ref{tab:parameter_charm2}.
  We also try to fit the data with the Gaussian form that is often employed in the NRp models, 
   however it provides an unreasonable $\chi^2/\text{d.o.f.}$ value.
  Note that we here use only the on-axis data which are expected to 
  less suffer from both the rotational symmetry breaking and discretization error, 
  because fit results obtained in each direction 
  significantly disagree with each other.
  We need the finer lattice to make a solid conclusion regarding the shape of the spin-spin potential 
  and also systematic uncertainties due to the rotational symmetry breaking.

\section{Nonrelativistic potential model with lattice inputs}
Using the quark kinetic  mass and 
the charmonium potentials determined by first principles of QCD,
we can solve the nonrelativistic Schr\"odinger equation for the bound $c\bar{c}$ systems
as same as calculations in the NRp models. 
In the BS amplitude method, a value of the difference $V_0 - E_{\rm ave}$
is directly obtained as the constant term in spin-independent charmonium potential,
while the value of $E_{\rm ave}$ is calculated through $E_{\rm ave} = M_{\rm ave}- 2m_Q$.
However statistical uncertainty of $m_Q$ is somewhat large
compared to an error of $V_0 - E_{\rm ave}$:
here these are $E_{\rm ave} = 0.508(69)$~GeV, $m_Q = 1.789(34)$~GeV and
$V_0 - E_{\rm ave} = -0.146(13)$~GeV.
To reduce statistical uncertainties,
we therefore solve the following  Schr\"odinger equation shifted by a constant energy~$-E_{\rm ave}$:
\begin{equation}
 \left\{  -\frac{1}{m_Q}\frac{\partial^2}{\partial r^2} + \frac{L(L+1)}{m_Qr^2} + V'_{SLJ}(r)
 \right\} u_{SLJ}(r) 
 =E'_{SLJ} u_{SLJ}(r) 
\label{eq_schre}
\end{equation}
where $V'_{SLJ}(r) = V_{SLJ}(r) - E_{\rm ave}$ and $E'_{SLJ} = E_{SLJ} - E_{\rm ave}$.
The interquark potential depends on the channel of charmonium states with $S$, $L$ and $J$.
Desired charmonium masses are obtained by merely adding $E'_{SLJ}$  to
the spin-averaged mass $M_{\rm ave}$ which is obtained from the standard lattice spectroscopy
with high accuracy:~$ M_{SLJ} = M_{\rm ave}+E'_{SLJ} = 2m_q  + E_{SLJ}$.

The resulting  potentials from lattice QCD are discretized in space~\cite{Charron:2013paa}. 
Therefore, instead of solving {\it continuum-type} Schr\"odinger equation, 
we practically solve eigenvalue problems as
\begin{eqnarray}
  \sum_{n=1}^{N_s/2-1} H_{m,n} u_n = E u_m \label{eq_eigen}
\end{eqnarray}
with a symmetric matrix defined in one of three specific directions
\begin{eqnarray}
  H_{n,n}      &=&  \frac{1}{\tilde{a}^2\mQ}\left[2 + \frac{L(L+1)}{n^2}  \right] + V'(n\tilde{a})  \\
  H_{n\pm 1,n} &=& - \frac{1}{\tilde{a}^2\mQ}.
\end{eqnarray}
The boundary condition to the reduced wave functions $u_n = u(n\tilde{a}) $ is simply set to $u_0 = 0$ and $u_{N_s/2} = 0$.
In this work, we separately solve Eq.~(\ref{eq_eigen}) in the directions of vectors $\bf r$ 
which are multiples of $(1, 0, 0)$, $(1, 1, 0)$ and $(1, 1, 1)$.
We prefer to use mainly on-axis data which is expected to receive smallest discretization errors and systematic uncertainties
due to rotational symmetry breaking,
and quote the largest difference between  on-axis and off-axis results
as the systematic error due to the choice of direction.
While statistical errors are estimated by the jackknife method.
A systematic uncertainty stemming from the choice of time window is relatively small
 compared with other errors.
Alternatively we can solve the Schr\"odinger equation in continuum space with the parameterized 
charmonium potential by empirical functional forms.
This procedure however highly depends on choice of functional forms especially at short distances, 
and give additional systematic uncertainties to resultant spectrum.

\begin{figure*}
  \includegraphics[width=.94\textwidth]{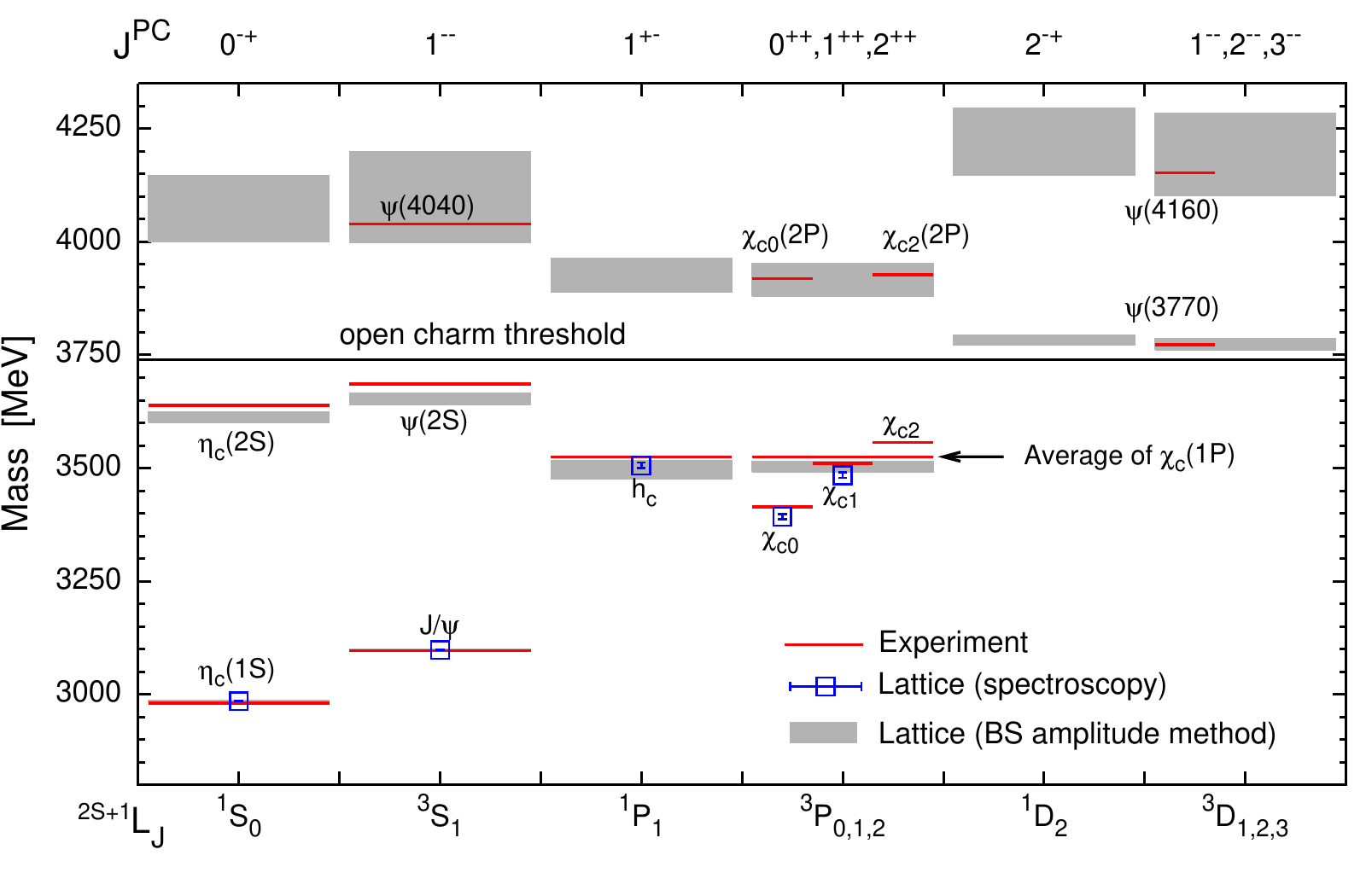}
  \centering
  \caption{Mass spectrum of charmonium states below and near the  open-charm threshold.
    The vertical scale is in units of MeV.
    Labels of $^{2S+1}[L]_J$ ($J^{PC}$) are displayed in lower (upper) horizontal axis.
    Rectangular boxes indicate predictions from the NRp models with theoretical inputs based on lattice QCD
    and their errors which are the sum of the statistical and systematic added in quadrature.
    Solid lines indicate  experimental values of well established charmonium states, while
    square symbols represent results of the standard lattice spectroscopy.
    A horizontal line shows the open-charm threshold.
    A symbol of $\ovli{^3P_J}$ denotes the spin-weighted average of spin-triplet $^3P_J$ states
    as $M_{\ovli{\chi_{cJ}}} = (M_{\chi_{c1}} +3M_{\chi_{c2}} + 5M_{\chi_{c2}})/9$.}
  \label{fig:potential_spec}
\end{figure*}

 \begin{table}
\centering
   \caption{
    Charmonium  mass spectrum is summarized in units of MeV.
    The labels of AVE and HYP in a column of ``state''  for $S$-states  denotes the spin-averaged mass 
    $(M_{^1S_0}+3M_{^3S_1})/4$ and hyperfine splitting mass $M_{^1S_0}-M_{^3S_1}$.   
    Experimental data~(denoted as Exp.) are taken from Particle Data Group,  rounded to $1$~MeV~\cite{Beringer:1900zz}.
    There are two lattice QCD results.
    First one is given by the usual spectroscopy, and 
    second one is a result calculated by 
    solving the Schr\"odinger equation with the charmonium potentials  determined from lattice QCD.
    For the second, first error is statistical, and second error is systematic error due to rotational symmetry breaking.
    For spin triplet states $^3[L]_J$, the spin-weighted average~($\ovli{^3[L]_J}$)  are also included.
   }
  \label{tab:higher_charmonium}
     \begin{tabular}{rcccc} \hline
       state &Exp. 
      & \multicolumn{2}{c}{Lattice QCD} & NRp model \\ 
        & & spectroscopy & BS amplitude  & \\[4pt] \hline
$\eta_c$         $(1^1S_0)$     & 2981 & 2985(1)   & 2985(2)(1)  & 2982 \\
$J/\psi$         $(1^3S_1)$     & 3097 & 3099(1)   & 3099(2)(1)  & 3090 \\
AVE                             & 3068 & 3070(9)   & 3070(2)(1)  & 3063 \\
HYP                             & 116  & 114(1)    & 113(1)(0)   & 108  \\[2pt]
		     	            		       	    
$\eta_c$         $(2^1S_0)$     & 3639 &             & 3612(9)(7)  & 3630 \\
$\psi$           $(2^3S_1)$     & 3686 &             & 3653(12)(5) & 3672 \\
AVE                             & 3674 &             & 3643(11)(5) & 3662 \\
HYP                             & 47   &             & 41(6)(3)    & 42   \\[2pt]
		     	            		       	    
$\eta_c$         $(3^1S_0)$     &      &             & 4074(20)(70)& 4043 \\
$\psi$           $(3^3S_1)$     & 4039 &             & 4099(24)(98)& 4072 \\
AVE                             &      &             & 4092(22)(91)& 4065 \\
HYP                             &      &             & 25(15)(28)  & 29   \\[2pt] \hline

$h_c$            $(1^1P_1)$     & 3525 & 3506(6) & 3496(7)(19) & 3516\\
$\ovli{\chi_{cJ}}$ $(\ovli{1^3P_J})$& 3525 &         & 3503(7)(10) & 3524\\
$\chi_{c0}$      $(1^3P_0)$     & 3415 & 3393(6) &             & 3424\\
$\chi_{c1}$      $(1^3P_1)$     & 3511 & 3485(6) &             & 3505\\
$\chi_{c2}$      $(1^3P_2)$     & 3556 &             &             & 3556\\[2pt]
		     	            
$h_c$            $(2^1P_1)$     &      &             & 3927(16)(34) & 3934\\
$\ovli{\chi_{cJ}}$ $(\ovli{2^3P_J})$&      &         & 3916(19)(31) & 3943\\
$\chi_{c0}$      $(2^3P_0)$     & 3918 &             &              & 3852\\
$\chi_{c1}$      $(2^3P_1)$     &      &             &              & 3925\\
$\chi_{c2}$      $(2^3P_2)$     & 3927 &             &              & 3972\\[2pt] \hline
$\eta_{c2}$      $(1^1D_2)$     &      &             & 3783(12)(4)  & 3799   \\
$\ovli{\psi}$      $(\ovli{1^3D_J})$&      &         & 3774(13)(2)  & 3800   \\
$\psi$           $(1^3D_1)$     & 3773 &             &              & 3785   \\
$\psi$           $(1^3D_2)$     &      &             &              & 3800   \\
$\psi$           $(1^3D_3)$     &      &             &              & 3806   \\[2pt]
		     	            
$\eta_{c2}$      $(2^1D_2)$     &      &             & 4221(21)(72) & 4158   \\
$\ovli{\psi}$      $(\ovli{2^3D_J})$&      &         & 4193(25)(88) & 4159   \\
$\psi$           $(2^3D_1)$     & 4153 &             &              & 4142   \\
$\psi$           $(2^3D_2)$     &      &             &              & 4158   \\
$\psi$           $(2^3D_3)$     &      &             &              & 4167  \\  \hline
     \end{tabular}
 \end{table}

Fig.~\ref{fig:potential_spec} shows the mass spectrum of the charmonia below 4200~MeV.
Theoretical spectrum plotted as rectangular boxes are given
by solving the discrete nonrelativistic Schr\"odinger
equations with theoretical inputs.
Quoted errors of charmonium masses are statistical and systematic uncertainties combined in quadrature.
For purpose of comparison, both experimental values and results of
the standard lattice spectroscopy are plotted together.
The experimental values  are taken from Particle Data Group~\cite{Beringer:1900zz}.
At first glance, 
we find that theoretical calculations from the NRp model with lattice inputs are in fairly good agreement 
with not only the lattice spectroscopy, but also experiments below open charm threshold.
All results are also summarized in Table~\ref{tab:higher_charmonium}.
In this study, we succeed in extracting only the spin-spin potential
among spin-dependent interquark potentials.
Thus at this stage we cannot predict the spin-orbit splitting
which is led by the tensor and spin-orbit forces.
In other wards, we can compute only the spin-averaged mass
for  excited states with higher angular momentum such as
$\chi_{cJ}$ state.

Our theoretical calculations for charmonium states
below the open-charm threshold are in fairly  good agreement with the experimental measurements.
The point we wish to emphasize here is that our novel approach has
no free parameters in solving the Schr\"odinger equation opposed to the phenomenological NRp model.
All of the parameters appeared in the NRp model calculation are solely determined by lattice QCD simulations, where
three light hadron masses ($M_\pi$, $M_K$ and $M_\Omega$) are used for inputs to fix
the lattice spacing and light quark hopping parameters.
Only experimental values of $\eta_c$ and $J/\psi$ masses in the charm sector are used to
determine the charm quark parameters  in the RHQ action.
In this sense the new approach proposed here is distinctly different from
the existing calculations with the phenomenological quark potential models.

 %
 %
\section{\label{summary}Summary}
We have calculated the interquark potentials between charm quark and anti-charm quark 
almost on the physical point.
The interquark potential at finite quark mass is defined through the equal-time Bethe-Salpeter wave function.
Our simulations have been performed in the vicinity of the physical light quark masses, 
which corresponds to $M_\pi = 156$~MeV,
using the PACS-CS gauge configurations generated with the Iwasaki gauge action and 2+1 flavors of Wilson clover quark.
We use the relativistic charm quark tuned to reproduce the experimental values of $J/\psi$ and $\eta_c$ masses.
The resulting spin-independent potential shows behavior of  Coulomb plus linear form, and their parameters are 
close to values used in the traditional quark potential models.
Also the string breaking due to existence of  sea quarks is not observed.
On the other hand, the spin-spin potential obtained from the dynamical simulations exhibits 
the short-range repulsive interaction. 
Its shape is quite different from the a repulsive $\delta$-function
potential induced by the one-gluon exchange  which are usually adopted in the quark potential model.

We have calculated the charmonium spectrum 
by solving nonrelativistic Schr\"odinger equation 
with the theoretical input of the spin-independent and spin-spin potentials and the quark kinetic mass. 
We simply solved the Schr\"odinger equation with Dirichlet boundary condition in a matrix manner. This approach enable us to directly use the raw
data of the charmonium potential without introducing a phenomenological parameterization for the discretized potential data.
We found an excellent agreement of low-lying charmonium masses between our results and the experimental data.
We emphasize that our novel approach has no free parameters in solving the Schr\"odinger type equation opposed to
conventional phenomenological quark potential models.
As for inputs of lattice QCD, we essentially use three light hadron masses ($M_\pi$, $M_K$ and $M_\Omega$)
for fixing the lattice spacing and light quark hopping parameters,
and two charmonium masses ($M_{\eta_c}$ and $M_{J/\psi}$) for determining the parameters of RHQ action. 

In order to precisely predict the mass spectrum above the open charm threshold, 
we must take into account the effects of not only the mass shift caused by mixing the $\QQbar$ states 
with $D\bar{D}$ continuum, but also $S$-$D$ mixing due to existence of the tensor force.
However, in this work, we simply ignore these effects and also
apply our new approach to the charmonium states above the open-charm threshold. 
The theoretical prediction of the nonrelativistic potential model with lattice inputs is basically 
consistent with the existing experimental data, 
although the systematic uncertainties due to the rotational symmetry breaking are rather large.
For more comprehensive prediction including spin-orbit splittings, however,
we must calculate all spin-dependent terms (spin-spin, tensor and spin-orbit forces).
Especially the tensor force introducing the $S$-$D$ mixing would shift even the masses of $1S$-states.
Also the larger spatial extent is required to address the systematic uncertainties due to the finite size effect
for the higher excited state that are supposed to possess wider wavefunction. 
%


 \begin{theacknowledgments}
   We would like to thank T. Hatsuda for helpful suggestions, H. Iida and Y. Ikeda for fruitful discussions. 
 This work was partially supported by JSPS/MEXT Grants-in- Aid (No. 22-7653, No. 19540265, and No. 21105504).
 T. Kawanai was partially supported by JSPS Strategic Young Researcher Overseas Visits Program
  for Accelerating Brain Circulation (No.R2411).
 \end{theacknowledgments}



\bibliographystyle{aipproc}   

\bibliography{template-8s}

\IfFileExists{\jobname.bbl}{}
 {\typeout{}
  \typeout{******************************************}
  \typeout{** Please run "bibtex \jobname" to optain}
  \typeout{** the bibliography and then re-run LaTeX}
  \typeout{** twice to fix the references!}
  \typeout{******************************************}
  \typeout{}
 }

\end{document}


\endinput